\documentclass[preprint]{emulateapj}

\shorttitle{Enhanced Radiative Recombination Continua of Fe and Ca}
\shortauthors{Ohnishi et al.}

\begin{document}
\title{Discovery of Enhanced Radiative Recombination Continua of He-like Iron and Calcium from IC\,443 and Its Implications}

\author{Takao~Ohnishi,\altaffilmark{1} Hiroyuki~Uchida,\altaffilmark{1} Takeshi~Go~Tsuru,\altaffilmark{1} Katsuji~Koyama,\altaffilmark{1,2} Kuniaki~Masai,\altaffilmark{3} and Makoto~Sawada\altaffilmark{4}}
\email{ohnishi@cr.scphys.kyoto-u.ac.jp}

\altaffiltext{1}{Division of Physics and Astronomy, Graduate School of Science, Kyoto University, Kitashirakawa-oiwake-cho, Sakyo-ku, Kyoto 606-8502, Japan}
\altaffiltext{2}{Department of Earth and Space Science, Graduate School of Science, Osaka University, 1-1 Machikaneyama, Toyonaka, Osaka 560-0043, Japan}
\altaffiltext{3}{Department of Physics, Tokyo Metropolitan University, 1-1 Minami-Osawa, Hachioji, Tokyo 192-0397, Japan}
\altaffiltext{4}{Department of Physics and Mathematics, Aoyama Gakuin University, Fuchinobe 5-10-1, Sagamihara, Kanagawa 252-5258, Japan}

\begin{abstract}
We present deep observations of the Galactic supernova remnant IC\,443 with the {\it Suzaku X-ray satellite}.
We find prominent K-shell lines from iron and nickel, together with a triangle residual at 8--10~keV, which corresponds to the energy of the radiative recombination continuum (RRC) of He-like iron.
In addition, the wavy residuals have been seen at $\sim$5.1 and $\sim$5.5~keV.
We confirm that the residuals show the first enhanced RRCs of He- and H-like calcium found in supernova remnants.
These facts provide robust evidence for the recombining plasma.
We reproduce the plasma in the 3.7--10~keV band using a recombining plasma model at the electron temperature 0.65~keV.
The recombination parameter $n_{\rm e}t$ ($n_{\rm e}$ is electron density and $t$ is elapsed time after formation of a recombining plasma) and abundances of iron and nickel are strongly correlated, and hence the errors are large.
On the other hand, the ratio of nickel to iron relative to the solar abundances is well constrained to 11$^{+4}_{-3}$ (1$\sigma$).
A possibility is that the large abundance ratio is a result of an asymmetric explosion of the progenitor star.
\end{abstract}

\keywords{ISM: abundances --- ISM: individual objects (IC\,443) --- ISM: supernova remnants --- X-rays: ISM}

\section{Introduction}
Thermal X-rays from the supernova remnants (SNRs) are due to shock-heated plasma.
Electrons are first heated to high temperatures ($T_{\rm e}$), and then the atoms are gradually ionized.
Therefore, the plasmas in young and intermediate-aged SNRs are often described by a non-equilibrium ionization plasma (NEI), or more specifically an ionizing plasma (IP).
Electron temperature ($T_{\rm e}$) is higher than that predicted from the mean ionization states of elements (ionization temperature: $T_{\rm z}$).
As SNRs age, the ionization rate decreases and is balanced by the recombination rate; therefore, the plasma reaches collisional ionization equilibrium (CIE) with $T_{\rm e} = T_{\rm z}$.
The reverse case is a recombination-dominant process ($T_{\rm e} < T_{\rm z}$)---recombining plasma (RP).
Previous studies on thermal X-ray spectra in SNRs have used IP to describe young SNRs and CIE plasma to describe old SNRs.
This canonical scenario for SNR plasma evolution has been challenged by the recent {\it Suzaku} discoveries of RP from several mixed-morphology SNRs \citep{Rho1998}, IC\,443, W49B, G\,359.1$-$0.5, W28, W44, G\,346.3$-$0.2, and G348.5$+$0.1 \citep{Yamaguchi2009, Ozawa2009, Ohnishi2011, Sawada2012, Uchida2012, Yamauchi2013, Yamauchi2014}.
As evidence of the existence of RP, the authors cite the detection of enhanced radiative recombination continua (RRCs), X-ray emissions that are made when free electrons are directly recombined with atoms in a bound state (free-bound transition).

RRC is most conspicuous for the transition of free electrons to the ground state of either He- or H-like atoms.
Most RRCs discovered so far are those of He-like magnesium (Mg), silicon (Si), and sulfur (S) below $\sim$4~keV, a clouded energy band in which many emission lines from abundant elements overlap on the RRC structures.
Thus, the RRC signatures are only revealed as saw-teeth like residuals when fit to the IP or CIE model.
On the other hand, RRC structures should be more conspicuous in the energy band 4--10 keV, because the RRC structures---those of iron (Fe) and calcium (Ca)---are more sparsely spaced, with no overlap of emission lines from the relevant elements in this energy band.
However, no RRC structures of Ca and Fe have been reported---except Fe from W49B \citep{Ozawa2009}---mainly due to the limited statistics in the high-energy band.
Therefore, we conducted deep observations on IC\,443, the most robust RP SNR and only the RRCs of Mg, Si, and S but no other higher Z elements have been found \citep{Yamaguchi2009}.

IC\,443 (G\,189.1$+$3.0) is located on the Galactic anti-center at a distance of 1.5~kpc \citep{Welsh2003}.
The remnant is associated with a dense giant molecular cloud \citep{Cornett1977} near the Gem OB1 association \citep{Humphreys1978}.
Thus, IC\,443 is likely a remnant of a core-collapse supernova.
Using the {\it ASCA} satellite, \citet{Kawasaki2002} found that the K-shell intensity ratios of H-like Si and S relative to He-like Si and S were significantly higher than those expected in the CIE plasma of the electron temperature determined from the bremsstrahlung continuum.
They concluded that the plasma in IC\,443 was RP.
\citet{Yamaguchi2009} found RRCs of Mg, Si, and S and estimated that $kT_{\rm e}$ and $kT_{\rm z}$ were, respectively, $\sim$0.6 and $\sim$1.0--1.2~keV, confirming the existence of RP.
In this paper, we report further evidence of RP on the basis of the new discoveries of RRCs from Ca and Fe based on {\it Suzaku} deep observations.
We discuss the characteristic features of RP.

\section{Observation and Data Reduction}
We used the X-ray Imaging Spectrometer \citep[XIS;][]{Koyama2007} on board the {\it Suzaku} satellite \citep{Mitsuda2007} to perform multiple observations of IC\,443.
The first observation was made during the Announcement of Opportunity cycle~1 (AO1) phase and the results were reported by \citet{Yamaguchi2009}.
Then deep observations of $\sim$370~ks exposure were performed in the AO7 phase.
The log of these observations is given in Table~\ref{table_observation}.

XIS comprises four X-ray CCD cameras located at the foci of individual X-Ray Telescopes \citep{Serlemitsos2007}; three (XIS~0, 2, 3) are front-illuminated (FI), while the other (XIS~1) is back-illuminated (BI) CCDs.
XIS~2 has not been functioning since 2006 November, and one quadrant of XIS~0 has been unavailable since 2009 June, both possibly damaged by the micro-meteorites.
For data reduction and spectral analysis, we used HEAsoft version~6.12 and SPEX \citep{Kaastra1996} version~2.02.04, respectively.
XIS data were reprocessed using xispi software and the calibration database was updated in 2013 March.
After screening with standard criteria,\footnote{http://heasarc.gsfc.nasa.gov/docs/suzaku/processing/criteria\_xis.html} the total effective exposure time was 409.7~ks.

\begin{deluxetable}{lcccr@{.}l}
  \tablecaption{Observation Logs\label{table_observation}}
  \tablehead{
    \multicolumn{1}{c}{ID} & Cycle & Obs. Data & ($\alpha$, $\delta$)$_{{\rm J}2000.0}$ & \multicolumn{2}{c}{Exposure\tablenotemark{a}}
  }
  \startdata
  501006010 & AO1 & 2007 Mar 6 & (6$^{\rm h}$17$^{\rm m}$11$^{\rm s}$, +22\arcdeg46\arcmin32\arcsec) & 42 & 0~ks \\
  507015010 & AO7 & 2012 Sep 27 & (6$^{\rm h}$17$^{\rm m}$11$^{\rm s}$, +22\arcdeg45\arcmin12\arcsec) & 101 & 8~ks \\
  507015020 & AO7 & 2013 Mar 27 & (6$^{\rm h}$17$^{\rm m}$11$^{\rm s}$, +22\arcdeg45\arcmin12\arcsec) & 59 & 3~ks \\
  507015030 & AO7 & 2013 Mar 31 & (6$^{\rm h}$17$^{\rm m}$11$^{\rm s}$, +22\arcdeg45\arcmin12\arcsec) & 131 & 1~ks \\
  507015040 & AO7 & 2013 Apr 06 & (6$^{\rm h}$17$^{\rm m}$12$^{\rm s}$, +22\arcdeg44\arcmin52\arcsec) & 75 & 5~ks
  \enddata
  \tablenotetext{a}{Effective exposure of the screened XIS data.}
\end{deluxetable}

\section{Analysis}
\begin{figure}
  \begin{center}
    \includegraphics{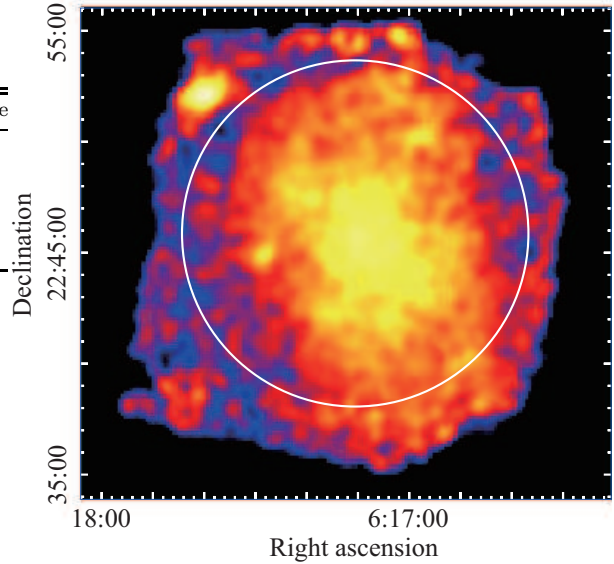}
    \caption{
      Vignetting-corrected XIS image of IC\,443 in the 3.7--10~keV band in logarithmic intensity scale.
      Data from XIS~0, 1, and 3 is combined.
      The white circle indicates the source region.
    }
    \label{figure_image}
  \end{center}
\end{figure}

Figure~\ref{figure_image} shows the vignetting-corrected XIS image in the 3.7--10~keV energy band after the subtraction of the non-X-ray background (NXB) constructed with the xisnxbgen software \citep{Tawa2008}.
This energy band includes the major spectral structures from argon (Ar) to nickel (Ni).
We make spectra from the solid circle of 8$'$ radius, excluding the regions of calibration sources (Figure~\ref{figure_image}).
We then subtract the NXB, which is made using all night-Earth observational data.
The spectra of the two FI CCDs (XIS~0 and 3) are merged to improve the statistics, because their response functions are almost identical.
We analyze only the merged FI spectrum because the NXB of the BI in the 6--10~keV band is higher than that of the source counts; therefore, NXB subtraction significantly degrades the source spectrum of BI.

\begin{figure} 
  \begin{center}
    \includegraphics{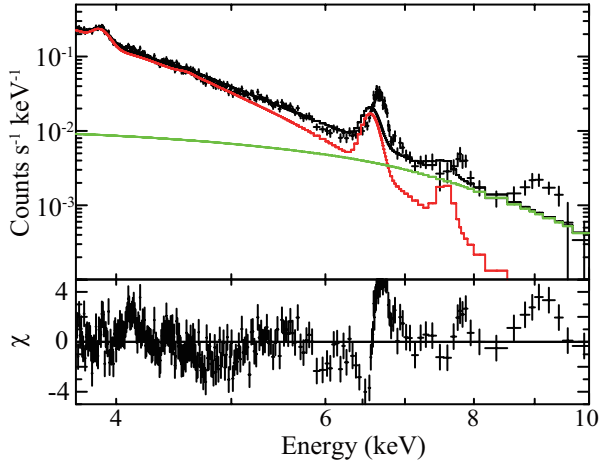}
    \caption{
      XIS spectrum in the 3.7--10~keV energy band.
      The best-fit CIE model is shown by the black solid line.
      The CIE and CXB components are given by the red and green lines, respectively.
      The lower panel shows the residuals from the best-fit model.
    }
    \label{figure_CIE}
  \end{center}
\end{figure}

Figure~\ref{figure_CIE} shows the merged FI spectrum in the 3.7--10~keV band.
We see prominent lines at 6.7 and~7.8 keV, which are, respectively, He$\alpha$ of Fe and He$\beta$ of Fe plus He$\alpha$ of Ni.
Here we define the transition lines from the first excited to the ground states (the principle quantum number $n=2\rightarrow1$) in He- and H-like atoms as He$\alpha$ and Ly$\alpha$, respectively, and the transition line from the second excited to the ground state ($n=3\rightarrow1$) in He-like atoms as He$\beta$.
In addition, a triangle residual at 8--10~keV corresponds to the energy of RRC of He-like Fe \citep[see e.g.,][]{Ozawa2009}.
The Ni line and RRC of Fe are the first detection with our deep observations.
We fit the spectrum with a model of optically thin thermal plasma in CIE.
The abundances of Ar, Ca, Fe, and Ni are free parameters, while those of the other elements are fixed to the solar abundances of \citet{Lodders2009}.
The cosmic X-ray background (CXB) spectrum is approximated by a power-law model with a photon index $\Gamma=1.412$ and the surface brightness in the 2--10~keV band of $6.38\times10^{-8}$~erg~cm$^{-2}$~s$^{-1}$~sr$^{-1}$ \citep{Kushino2002}.
Because IC\,443 is located in the Galactic anticenter direction, the contribution of Galactic ridge X-ray emission is negligible.
This CIE fit is completely rejected with the large $\chi^2$/degrees of freedom (dof) of 1080/271 ($=3.98$).

As is shown in Figure~\ref{figure_CIE}, the CIE model fails to reproduce the energies of the He$\alpha$ of Fe and Ni at $\sim$6.7 and $\sim$7.8~keV, respectively.
The line-like residuals at $\sim$7.0 and $\sim$4.1~keV are the Ly$\alpha$ of H-like Fe and Ca, respectively.
The other large residual is a triangle-shaped structure at 8--10~keV.
Considering the energy resolution of approximately 300~eV (at 8--10~keV), the energy of leading edge of the triangle is $\sim$8.9 keV, which corresponds to the binding energy of the ground state of He-like Fe, and thus the triangle would be an RRC of He-like Fe.

\begin{figure} 
  \begin{center}
    \includegraphics{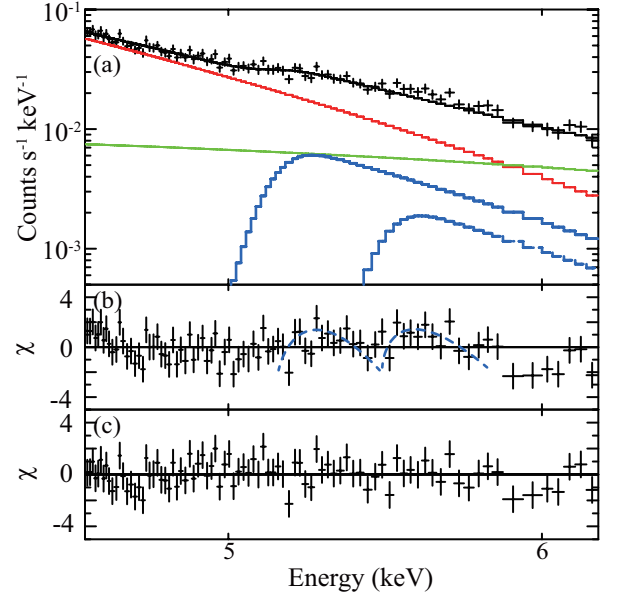}
    \caption{
      Uppermost panel (a) shows XIS spectrum in the 4.6--6.2~keV energy band with the best-fit bremsstrahlung emission (red), RRCs from He- and H-like Ca (blue), and CXB (green).
      Middle panel (b) represents the residuals from the models of bremsstrahlung emission and CXB.
      Lowermost panel (c) is the same as (b), except that it is taken from the RRCs of He- and H-like Ca.
      Blue dotted line in panel (b) is to guide the eye.
    }
    \label{figure_CaRRC}
  \end{center}
\end{figure}

In Figure~\ref{figure_CIE}, we also see a wavy residual at 5--6 keV.
The leading edge energies of the wave are $\sim$5.1 and $\sim$5.5~keV, which are likely the RRCs of He- and H-like Ca.
To examine these possible RRCs of Ca, we magnify the spectrum in the 4.6--6.2~keV band and fit it with a model of bremsstrahlung emission plus fixed CXB.
This model is not accepted with the $\chi^2$/dof $=113/85$ ($=1.33$), leaving a wavy excess at 5--6~keV, as shown in Figure~\ref{figure_CaRRC}(b).
We then add two RRC models for this wavy structure with the leading edges at 5.13 and 5.48~keV, which are the predicted values of He- and H-like Ca RRCs, respectively.
For the RRC model, an exponential decay tail of an {\it e}-folding temperature is added above the edge energy.
The temperature for the RRC and that of the bremsstrahlung are free parameters but are linked with each other.
This RRC model significantly reduces $\chi^2$/dof to 82/83 ($=0.99$), which yields an $F$-test probability of $2\times10^{-6}$.
The best-fit result and parameters are shown in Figure~\ref{figure_CaRRC}(a) and (c) and in Table~\ref{table_CaRRC}.
Thus, in the spectrum of IC\,443, we confirm for the first time the presence of enhanced RRCs of Ca.
Taken together, these facts provide good evidence for a RP.

\begin{deluxetable}{ccr@{}l}
  \tablecaption{Best-fit Parameters of RRC Model\label{table_CaRRC}}
  \tablehead{
    Component & Parameter & \multicolumn{2}{c}{Value\tablenotemark{a}}
  }
  \startdata
  Bremsstrahlung & $kT_{\rm e}$~(keV) & 0. &627$^{+0.060}_{-0.070}$ \\
  RRC of He-like Ca & Edge~(keV) & 5. & 13~(fixed) \\
  & VEM~(10$^{51}$~cm$^{-3}$)\tablenotemark{b} & 1. & 47$^{+0.23}_{-0.27}$ \\
  RRC of H-like Ca & Edge~(keV) & 5. & 48~(fixed) \\
  & VEM~(10$^{50}$~cm$^{-3}$)\tablenotemark{b} & 4. & 44$^{+2.09}_{-2.38}$ \\
  \hline
  $\chi^2$/dof & & 82 & /83 (0.99)
  \enddata
  \tablenotetext{a}{The errors are at 1$\sigma$.}
  \tablenotetext{b}{Volume emission measure at the distance 1.5~kpc.
    The units is ${\int}n_{\rm e}n_{\rm z}dV$, where $n_{\rm e}$, $n_{\rm z}$, and $V$ are the electron and Ca ions density~(cm$^{-3}$), and the emitting volume~(cm$^3$), respectively.}
\end{deluxetable}

\begin{figure}  
  \begin{center}
    \includegraphics{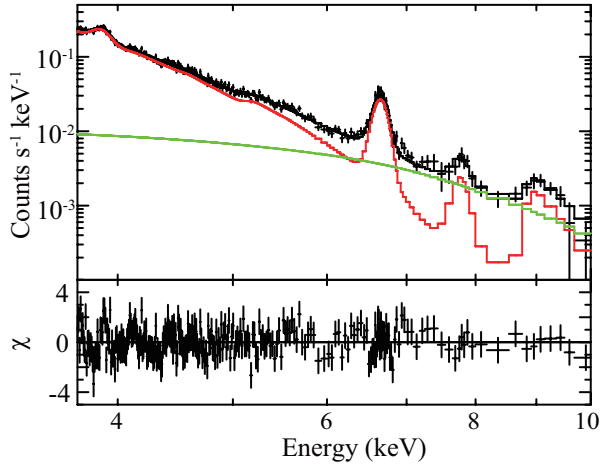}
    \caption{
      Same as Figure~\ref{figure_CIE}, but with the best-fit NEIJ model.
    }
    \label{figure_NEIJ}
  \end{center}
\end{figure}

Therefore, we fit the 3.7--10~keV spectrum with an NEIJ model \citep{Kaastra1996, Yamaguchi2012, Sawada2012, Uchida2012}, which represents a recombination-dominant phase starting from an initial ionization temperature $kT_{{\rm z}0}$ and an electron temperature $kT_{\rm e}$, after the recombination time scale of $n_{\rm e}t$, where $n_{\rm e}$ and $t$ are the density of electrons and elapsed time, respectively.
In the fitting, $kT_{{\rm z}0}$ cannot be constrained, with a lower limit of $\sim$3~keV. We therefore fit the spectrum by fixing $kT_{{\rm z}0}$ to 10~keV.
The other parameters---normalization, $kT_{\rm e}$, $n_{\rm e}t$, and abundances of Ar, Ca, Fe and Ni---are free parameters.
This NEIJ model is acceptable with $\chi^2$/d.o.f. $=302$/270 ($=1.12$).
The best-fit model is given in Figure~\ref{figure_NEIJ} and the parameters in Table~\ref{table_NEIJ}.
Because the CXB intensity has a fluctuation 6.49\% (1$\sigma$) in a $\sim$0.5~deg$^2$ area \citep{Kushino2002}, which corresponds to $\sim$20\% in the region of the SNR spectrum, we check for possible systematic error due to this CXB fluctuation changing the intensity of the CXB by $\pm$20\%.
We find no significant change of the best-fit results, except some incrementing of the error ranges (systematic errors).

\begin{deluxetable}{cr@{}l}
  \tablecaption{Best-fit Parameters of the NEIJ Model\label{table_NEIJ}}
  \tablehead{
  Parameter & \multicolumn{2}{c}{Value\tablenotemark{a}}
  }
  \startdata
  $kT_{\rm e}$~(keV) & 0. & 65$\pm0.02\pm0.02$ \\
  $kT_{{\rm z}0}$~(keV) & 10 & ~(fixed) \\
  $n_{\rm e}t$~($10^{11}$~cm$^{-3}$~s) & 9. & 8$\pm0.6^{+0.5}_{-0.6}$ \\
  $Z_{\rm Ar}$~(solar) & 0. & 76$^{+0.46}_{-0.32}$$^{+0.70}_{-0.17}$ \\
  $Z_{\rm Ca}$~(solar) & 1. & 7$^{+0.5}_{-0.3}$$^{+0.9}_{-0.2}$ \\
  $Z_{\rm Fe}$~(solar) & 2. & 7$^{+1.6}_{-1.0}$$^{+4.5}_{-0.6}$ \\
  $Z_{\rm Ni}$~(solar) & 30 & $^{+26}_{-14}$$^{+57}_{-5}$ \\
  VEM\tablenotemark{b}~($10^{57}$~cm$^{-3}$) & 6. & 7$^{+0.8}_{-1.2}$$^{+0.4}_{-0.9}$ \\
  \hline
  $\chi^2$/dof &  302 & /270 (1.12)
  \enddata
  \tablenotetext{a}{The errors are at 1$\sigma$ of statistical and systematic.}
  \tablenotetext{b}{Volume emission measure at the distance of 1.5~kpc.
    The units is ${\int}n_{\rm e}n_{\rm p}dV$, where $n_{\rm p}$ is proton density~(cm$^{-3}$).}
\end{deluxetable}

\section{Discussion}
Based on deep observations, we discover a strong He-$\alpha$ of Fe at $\sim$6.7~keV.
Because the electron temperature is very low---at most $\sim$0.7~keV---it is almost impossible to emit He-$\alpha$ of Fe by collisional excitation, a dominate process to produce this line in IP or CIE plasma.
Therefore, the presence of this line itself already indicates that the plasma is neither IP nor CIE.
Furthermore, we discover RRCs of He-like Fe and Ca and H-like Ca, which is robust evidence for RP.
The RRC of He-like Fe is the second sample after W49B, while the RRCs of He- and H-like Ca are the first discoveries in SNRs.

Although the line flux of Fe He-$\alpha$ has sufficient statistics, the best-fit 1$\sigma$ error of the Fe abundance is significantly large.
This situation is somewhat similar to those of Ca and Ni.
This apparent ``inconsistency'' is due to the RP proper characteristic, and thus is not found in IP.

\begin{figure}
  \begin{center}
    \includegraphics{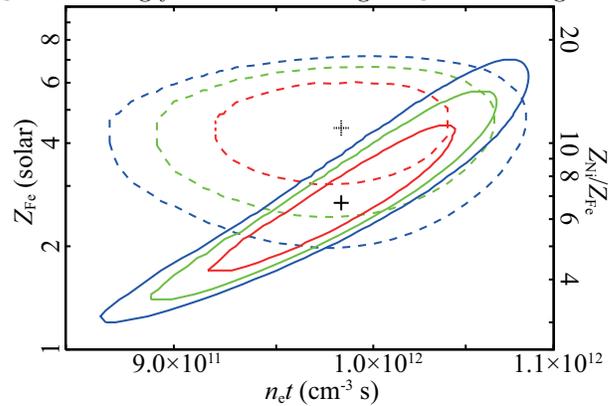}
    \caption{
      Solid lines show the error contours for the Fe abundance--$n_{\rm e}t$ space (left {\it y}-axis) and dashed lines show the abundance ratio of $Z_{\rm Ni}$/$Z_{\rm Fe}$--$n_{\rm e}t$ space (right {\it y}-axis), both in double-logarithmic scale.
      Confidence levels are at 1$\sigma$ (red), 2$\sigma$ (green), and 3$\sigma$ (blue).
      The crosses show the best-fit parameters
    }
    \label{figure_net_Fe}
  \end{center}
\end{figure}

The ionization rate from less ionized atoms is higher than that from more ionized atoms, and thus the number of charges in IP increases monotonically as $n_{\rm e}t$ increases.
At any $n_{\rm e}t$, no large dispersion of the charge number appears.
Unlike the ionization rate, the recombination rate does not strongly depend on the charge number, and thus a broad distribution of different ions is achieved in RP at a large $n_{\rm e}t$.
In our case, H-like Fe can survive in a wide range of $n_{\rm e}t$, significantly varying from 10$^{11}$ to 10$^{12}$~cm$^{-3}$~s.
Near $\sim$10$^{12}$~cm$^{-3}$~s, the fraction of H-like Fe decreases to almost 0\%.
RRCs of He-like atoms occur due to the recombination of free electrons to the ground states of H-like ions.
At low electron temperature (0.65~keV), collisional excitation rate of He-like atoms from ground to excited states is far smaller than that of the recombination process.
The contribution of the cascade line, which originates from electrons captured at the excited levels of ions by free-bound transition \citep{Ozawa2009}, is significantly larger than that of the line flux that originates from collisional excitation.
Therefore, the flux of He-like K-shell lines and RRC are approximately proportional to the fraction of H-like ions.
Accordingly, in RP, atomic abundances and $n_{\rm e}t$ are strongly correlated---larger $n_{\rm e}t$ leads to greater abundance.
We plot the confidence contour in the two-dimensional space of the relaxation parameter $n_{\rm e}t$ and the metal abundance of Fe (Figure~\ref{figure_net_Fe}).

The detection of the Ni He$\alpha$ line and a large ratio of $Z_{\rm Ni}$/$Z_{\rm Fe}$ relative to the solar abundances of $\sim$10 are also discoveries from IC\,443 and are shown in Figure~\ref{figure_net_Fe}.
To check this high ratio of $Z_{\rm Ni}$/$Z_{\rm Fe}$, we compare SPEX with the other sets of atomic data concerned \citep{Masai1997,Bryans2009} and estimate the uncertainty to be 20\%.
We thus conclude that the ratio of $Z_{\rm Ni}$/$Z_{\rm Fe}$ relative to the solar abundances is in the range of $\sim$8--12.
Unlike the absolute value of abundance for each element, this high ratio of $Z_{\rm Ni}$/$Z_{\rm Fe}$ is valid in the error range of $n_{\rm e}t$.
However, such a high ratio is not predictable from any theoretical model of spherically symmetric explosion in core-collapse supernovae \citep[e.g.,][]{Woosley1995}.
Recently, a similar high ratio of Ni to Fe relative to the solar abundances $\sim$8 was obtained from core-collapse SNRs, G\,350.1$-$0.3 and G\,349.7$+$0.2, suggesting that a significant fraction of Ni is ejected from the core region of their progenitors \citep{Yasumi2014}.
For SN\,2006aj, \citet{Maeda2007} proposed a model in which a large amount of $^{58}$Ni might be ejected from the core as a result of asymmetric explosion.
A possibility is that IC 443 is also a remnant of an asymmetric explosion of a core-collapse supernova.

\acknowledgements
The authors thank all Suzaku team members for developing hardware and software, operating the spacecraft, and calibrating instruments.
T.O. and H.U. are supported by the Japan Society for the Promotion of Science (JSPS) Research Fellowship for Young Scientists.
This work is supported by JSPS Scientific Research grant numbers 20340043, 23340047 and 25109004 (T.G.T.), 23000004 and 24540229 (K.K.), 22540253 (K.M.), and 24840036 (M.S.).

\end{document}